\documentclass[twocolumn]{aastex631}

\def\specialname[#1]{\textbf{\textsc{#1}}}
\usepackage{amsmath}

\newcommand{\sersic}{S\'{e}rsic}

\shorttitle{Revisiting the fundamental metallicity relation}
\shortauthors{Ma et al.}

\graphicspath{{./}{figures/}}

\begin{document}

\title{Revisiting the fundamental metallicity relation with observation and simulation}


\author[0009-0006-7343-8013]{Chengyu Ma}
\affiliation{Department of Astronomy, University of Science and Technology of China, Hefei 230026, China}
\affiliation{School of Astronomy and Space Science, University of Science and Technology of China, Hefei 230026, China}

\author[0000-0002-3775-0484]{Kai Wang}\thanks{E-mail: wkcosmology@gmail.com}
\affiliation{Kavli Institute for Astronomy and Astrophysics, Peking University, Beijing 100871, China}
\affiliation{Institute for Computational Cosmology, Department of Physics, Durham University, South Road, Durham, DH1 3LE, UK}
\affiliation{Centre for Extragalactic Astronomy, Department of Physics, Durham University, South Road, Durham DH1 3LE, UK}

\author[0000-0003-1588-9394]{Enci Wang}\thanks{E-mail: ecwang16@ustc.edu.cn}
\affiliation{Department of Astronomy, University of Science and Technology of China, Hefei 230026, China}
\affiliation{School of Astronomy and Space Science, University of Science and Technology of China, Hefei 230026, China}

\author{Yingjie Peng}
\affiliation{Kavli Institute for Astronomy and Astrophysics, Peking University, Beijing 100871, China}

\author[0009-0006-1483-4323]{Haochen Jiang}
\affiliation{Department of Astronomy, University of Science and Technology of China, Hefei 230026, China}

\author{Haoran Yu}
\affiliation{Department of Astronomy, University of Science and Technology of China, Hefei 230026, China}

\author{Cheng Jia}
\affiliation{Department of Astronomy, University of Science and Technology of China, Hefei 230026, China}

\author{Zeyu Chen}
\affiliation{Department of Astronomy, University of Science and Technology of China, Hefei 230026, China}

\author{Haixin Li}
\affiliation{Department of Astronomy, University of Science and Technology of China, Hefei 230026, China}

\author[0000-0002-7660-2273]{Xu Kong}
\affiliation{Department of Astronomy, University of Science and Technology of China, Hefei 230026, China}
\affiliation{School of Astronomy and Space Science, University of Science and Technology of China, Hefei 230026, China}

\begin{abstract}
    The gas-phase metallicity of galaxies is regulated by multiple
    astrophysical processes, which makes it a crucial diagnostic of galaxy
    formation and evolution. Beyond the fundamental mass-metallicity relation,
    a debate about the secondary galaxy property to predict the metallicity of
    galaxies arises. Motivated by this, we systematically examine the
    relationship between gas-phase metallicity and other galaxy properties,
    i.e. star formation rate (SFR) and galaxy size, in addition to stellar mass
    in both observation and simulation. We utilize the data from the MaNGA
    (Mapping Nearby Galaxies at Apache Point Observatory) survey and the TNG50
    simulations. We find that the combination of $M_*/R_{\rm e}^\beta$ with
    $\beta\sim 0.6-1$ is in much stronger correlation to the metallicity than
    stellar mass alone, regardless of whether the SFR is included or not, in
    both observation and simulation.  This indicates that galaxy size plays a
    more important role in determining gas-phase metallicity of galaxies than
    SFR.  In addition, the TNG simulation predicts that the SFR, although being
    a subdominant role, becomes increasingly important in high-$z$ universe.
    Finally, we speculate that SFR modulates metallicity on the temporal
    dimension, synchronized with time-varying gas inflows, and galaxy size
    regulates metallicity on the spatial dimension by affecting the
    gravitational potential and the mass loading factor.
\end{abstract}

\keywords{galaxies: general; galaxies: evolution; galaxies: statistics; galaxies: fundamental parameters}

\section{Introduction}%
\label{sec:introduction}

Galaxies formation and evolution are regulated by multiple astrophysical
processes, including gas cooling and accretion, star formation, stellar and AGN
feedback and associated gas outflows, the recycling of the ejected gas, and
others \citep[e.g.][]{Schaye-10, Bouche-10, Dave-11, Lilly-13, Peng-14,
    Belfiore-19, Wang-19, Wang-21, Wang-22,
wangEnvironmentalDependenceMassMetallicity2023}. Each of these processes
leaves their signature on the metal content of galaxies, which, in turn,
makes metallicity a crucial diagnostic of these physical processes during
galaxy formation and evolution. However, due to the complexity of the
situation, it is unpractical to directly infer the role played by each
individual physical process from the metal content of the galaxy. Instead,
we first establish the scaling relations between the metal content and
other galaxy properties, and comprehend these scaling relations with the
facility of semi-analytical models and numerical simulations, which has
already become a common practice.

The most fundamental scaling relation is the mass-metallicity relation
\citep[e.g.][]{Lequeux-79, Tremonti-04}, which says that gas in massive
galaxies is more metal-enriched than those in low-mass galaxies. Specifically,
based on large numbers of galaxies from the Sloan Digital Sky Survey
\citep[SDSS;][]{Stoughton-02}, \cite{Tremonti-04} found a strong correlation
between stellar mass and gas-phase metallicity for star-forming galaxies, which
is referred to as the mass-metallicity relation (MZR). MZR is also established
at high $z$ but the overall amplitude decreases with increasing redshift, which
indicates that high-$z$ galaxies are more metal-poor compared to their low-$z$
counterparts at fixed stellar mass \citep[e.g.][]{Savaglio-05, Maier-06,
Maiolino-08}. The correlation between stellar mass and gas-phase metallicity
can be driven by several factors, including outflows driven by supernova winds
\citep{Larson-74, Finlator-08, Bassini-24} and varying star formation
efficiencies in galaxies \citep{Brooks-07, Calura-09}.

Despite the strong correlation between gas-phase metallicity and stellar
mass, considerable scatter still remains. Further investigation found that the
residual gas-phase metallicity with respect to the mean relation is correlated
to the current star formation rate (SFR) of galaxies, and the joint correlation
among stellar mass, gas-phase metallicity, and SFR is known as the fundamental
metallicity relation \citep[FMR;][]{Mannucci-10, Lara-Lopez-10, Richard-11,
    Nakajima-12, Salim-14, Cresci-19, Huang-19, Curti-20, Curti-24, Garcia-24,
Perez-Diaz-24}. In particular, \cite{Mannucci-10} proposed a universal,
epoch-independent mass-metallicity-SFR relation. They suggested that the
apparent evolution in the MZR could be explained, phenomenologically, by the
redshift evolution of the star-forming main sequence. However, recent studies
based on deep JWST/NIRSpec spectroscopy found that high-$z$ galaxies,
especially $z>6$, are significantly metal-deficient compared with counterparts
in the local Universe with controlled stellar mass and SFR \citep{Curti-24,
Perez-Diaz-24}, which challenges the epoch-independency of FMR.

Recently, growing evidence shows that the gravitational potential of galaxies
plays a more fundamental role in regulating the metal content of star-forming
galaxies, rather than the stellar mass
\citep[e.g.][]{ellisonCluesOriginMassMetallicity2008, DEugenio-18,
    sanchezalmeidaOriginRelationMetallicity2018a,Huang-19,
    sanchez-menguianoStellarMassNot2024,
sanchez-menguianoMoreFundamentalFundamental2024, Ma-Du-24}. Similar conclusion
was drawn on the stellar metallicity of galaxies \citep{Barone-20, Vaughan-22,
Cappellari-23}. These results motivate us to take a closer look at the
statistical relationship between gas-phase metallicity and other galaxy
properties, which are stellar mass, SFR, and galaxy size in this study, for
star-forming galaxies in both observation and simulation, aiming to pin down
the fundamental determinant of the metallicity of the gas content in
star-forming galaxies. We find consistent results across observations and
simulations, which shows that stellar mass and galaxy size play primary roles,
while SFR plays a secondary role in determining the gas-phase metallicity of
star-forming galaxies.

This paper is structured as follows: \S\,\ref{sec:data} introduces both the
observational data and numerical simulations. \S\,\ref{sec:results} presents
our results obtained from aforementioned data. Finally,
\S\,\ref{sec:discussion_and_summary} presents the discussion about the
implications on galaxy formation and evolution, together with the summary to our
main findings.

\section{Data}%
\label{sec:data}

\subsection{Observational data}%
\label{sub:observational_data}

The observation sample which we select to research is from MaNGA \citep[Mapping
Nearby Galaxies at Apache Point Observatory;][]{Bundy-15} Data Release 17
\citep{Abdurrouf-22}. Using the two dual-channel BOSS spectrographs at the
Sloan Telescope \citep{Gunn-06, Smee-13}, MaNGA covers the wavelength of
3600–10300 \AA\ at resolution of $\sim$2000. The spatial coverage of individual
galaxies is typically larger than 1.5 Re with a spatial resolution of 1–2 kpc.

The measurements of stellar mass and total SFR are taken from
\cite{Salim-18}\footnote{https://salims.pages.iu.edu/gswlc/}, which are derived
from the spectral energy distribution fittings of GALEX, SDSS, and WISE
photometry. Based on these two parameters, we derive the star-forming main
sequence (SFMS) based on the iterative algorithm presented in
\citet{wangDissectTwohaloGalactic2023} \citep[see
also][]{wooDependenceGalaxyQuenching2013, donnariStarFormationActivity2019}. We
start from an initial guess of the linear function then iteratively select
galaxies within 1 dex of the SFMS and re-fit the slope and the intercept until
convergence. The result SFMS is
\begin{equation}
    \log \left(\frac{\rm SFR}{{\rm M_{\odot}yr^{-1}}}\right) = 0.71\times \log
    \left(\frac{M_*}{{\rm M_{\odot}}}\right) - 7.27.
\end{equation}
Anchored with this SFMS, we select 5532 star-forming galaxies that lie within 1
dex of this linear function for further analysis in this work.

Since \cite{DEugenio-18} found that it is critical to use the aperture-matched
metallicity in studying the metallicity scaling relation of galaxies, we in
this work use the H$\alpha$-luminosity weighted gas-phase metallicity measured
within effective radius ($R_{\rm e}$) as a representative of overall
metallicity of MaNGA galaxies \citep[also see][]{Wang-21}. The effective radius
is measured from the \sersic\ fitting on SDSS $r$-band image. The gas-phase
metallicity of MaNGA galaxies is computed with three methods using different
combination of strong lines, which are the {\tt N2S2H$\alpha$} diagnostic
introduced by \cite{Dopita-16}, the {\tt S-calibration} ({\tt Scal} for short)
estimator\citep{Pilyugin-16}, and the {\tt N2O2} diagnostic \citep{Dopita-13,
Zhang-17}. These three metallicity indicators are particularly adopted due to
the following reasons.

The {\tt N2S2H$\alpha$} is insensitive to reddening, and \cite{Easeman-24}
proposed that the {\tt N2S2H$\alpha$} is preferred when studying the
distribution of metals within galaxies, because {\tt N2S2H$\alpha$} shows a
near-linear relation with T$e$-based measurement. The {\tt Scal} indicator
leverages three emission line ratios, enhancing the accuracy over previous
strong-line methods. \cite{Pilyugin-16} demonstrated that the {\tt Scal}
indicator provides metallicity measurements that closely align with the $T_{\rm
e}$-based methods, exhibiting a scatter of only approximately 0.05 dex within
the metallicity range of $7.0 < 12 + \log({\rm O/H}) < 8.8$. {\tt N2O2} is not
sensitive to ionization parameter or ionizing spectrum hardness but to
metallicity \citep{Kewley-02, Dopita-13}, while it relies on the N/O. By
studying the metallicity of HII regions and diffuse ionized gas (DIG),
\cite{Zhang-17} proposed that {\tt N2O2} are optimal among many metallicity
indicators in case of the existence of DIG.

\subsection{Cosmological galaxy formation simulation}%
\label{sub:cosmological_galaxy_formation_simulation}

This work employs the state-of-the-art simulation suite of the Illustris The
Next Generation (TNG), which comprises several cosmological galaxy formation
simulations with different resolutions. Here we use the one with the highest
resolution: TNG50 \citep{Nelson-18, Pillepich-18}. The stellar mass, $M_*$, and
SFR are calculated within twice the effective radius, while the effective
radius, $R_{\rm e}$, is calibrated so that a sphere with this radius encloses
half of the total stellar mass in each subhalo. Here subhalos are identified
with the \texttt{SUBFIND} algorithm
\citep{springelSimulationsFormationEvolution2005} using all types of particles
in each FoF halo, which is identified using the conventional Friends-of-Friends
algorithm with only dark matter particles. The gas-phase metallicity, $12 +
\log[{\rm O/H}]$, is inferred as the ratio between the abundances of oxygen and
hydrogen.

For the sake of reliable gas-phase metallicity calculation, we only include
galaxies on the SFMS, which is determined using the iterative algorithm
presented in \citet{wangDissectTwohaloGalactic2023} and the result at $z=0$ is
\begin{equation}
    \log\left(\frac{\rm SFR}{\rm M_\odot yr^{-1}}\right) = 0.75\times
    \log\left(\frac{M_*}{\rm M_\odot}\right) - 7.72,
\end{equation}
and we only include galaxies with \texttt{SubhaloFlag==1}.

For a fair comparison between observation and simulation, we also need to
control the aperture within which all these physical parameters are measured.
For stellar mass and star formation rate, since current cosmological
hydrodynamical simulations cannot deliver physical values but rely on the
calibration against observational statistics, like stellar mass functions
\citep[see the introduction in][]{schayeEAGLEProjectSimulating2015}, we would
better use the aperture used for calibrating the simulation, which is $2R_{\rm e}$ in
all TNG simulations. The gas-phase metallicity is measured within $R_{\rm e}$
in observation for the sake of data quality, and we choose to use $2R_{\rm e}$
in simulation for physical consistency. In addition, we have checked the
results where the gas-phase metallicity was measured within $R_{\rm e}$ for all
simulated galaxies and it does not impact our scientific conclusions drawn
in this work.

\section{Results}%
\label{sec:results}

\subsection{Mass-metallicity relation and its relationship to SFR and size}%
\label{sub:mass_metallicity_relation_and_its_relationship_to_sfr_and_size}

\begin{figure*}
    \centering
    \includegraphics[width=\linewidth]{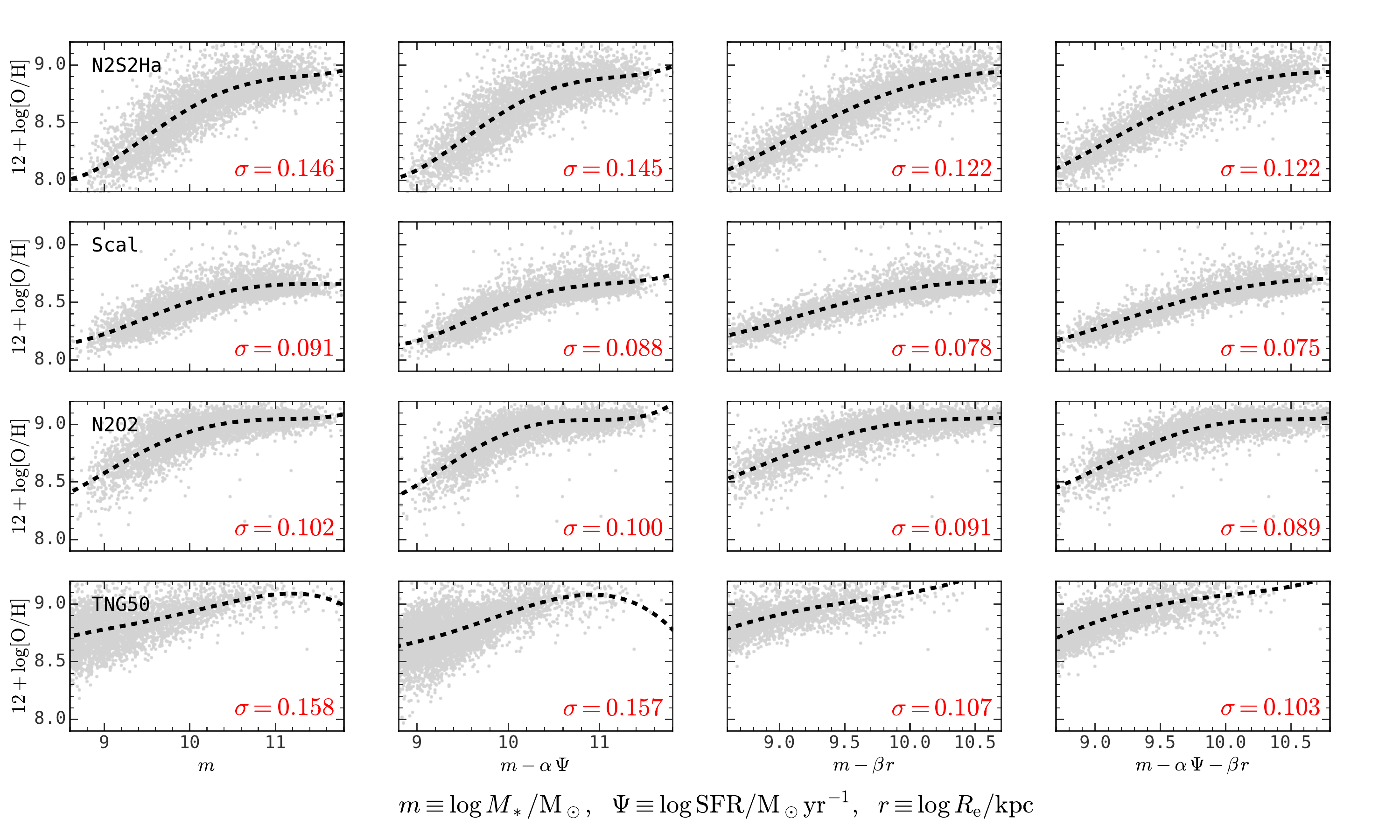}
    \caption{
        The mass-metallicity relation (first column) of observed galaxies, and
        its relation to SFR (second column), galaxy size (third
        column), and two quantities combined (forth column). Here three
        different gas-phase metallicity estimators are used ({\tt N2S2Ha}
        \citet{dopitaChemicalAbundancesHighredshift2016}, {\tt Scal}
        \citet{pilyuginNewCalibrationsAbundance2016}). The red text in each
        panel shows the standard deviation to the 4th-order fitting curve shown
        in black dashed line, which is denoted as $\sigma$. One can see that
        $\sigma$ is decreasing from left panels to right panels. Particularly,
        the panels of the third column exhibit lower $\sigma$ compared to those
        of the second panel, which indicates that galaxy size is a better
        predictor of the deviation in the mass-metallicity relation compared to SFR.
    }%
    \label{fig:figures/mzr}
\end{figure*}

\begin{figure}
    \centering
    \includegraphics[width=0.9\linewidth]{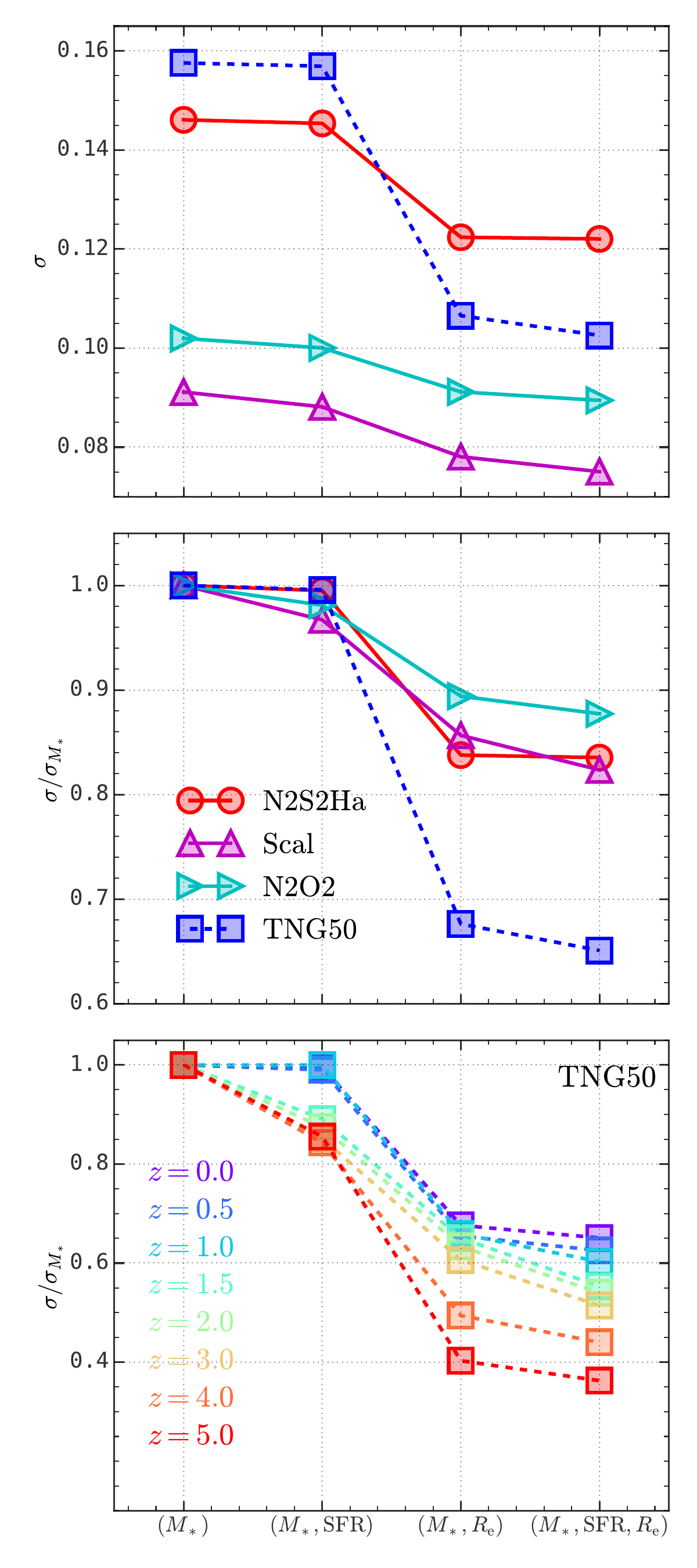}
    \caption{
        The standard deviation of the residual to the fitting curves in
        Fig.\,\ref{fig:figures/mzr} for four different gas-phase
        metallicity estimators. This figure shows that the inclusion of SFR
        only reduce $\sigma$ by $\lesssim 4\%$, but the inclusion of $R_{\rm
        e}$ can reduce $\sigma$ by $10-16\%$. Besides, including $R_{\rm e}$
        alone can obtain similar performance as including both $R_{\rm e}$ and
        SFR.
    }%
    \label{fig:figures/sigma}
\end{figure}

\begin{figure*}
    \centering
    \includegraphics[width=\linewidth]{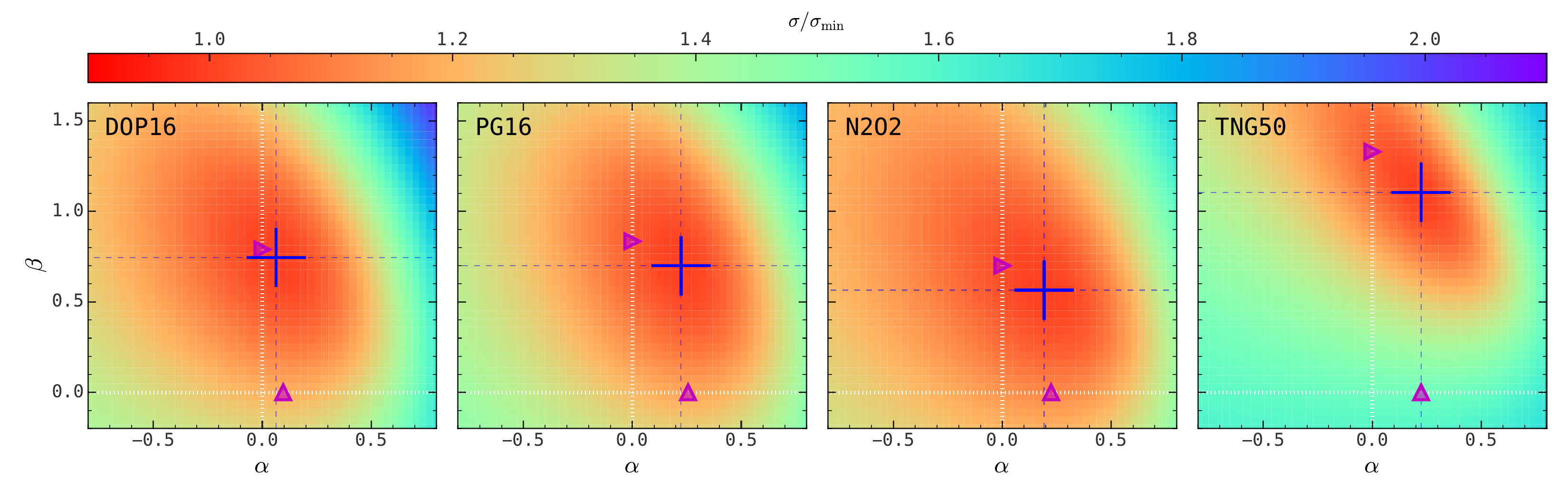}
    \caption{
        The scatter of the metallicity scaling relations in Figure
        \ref{fig:figures/mzr} as a function of the values of $\alpha$ and
        $\beta$ for the observations and TNG50. In each panel, the magenta
        triangles show the values of $\alpha$ and $\beta$ when fitted
        individually with the other parameter sets to zero, and the blue
        crosses show the results when fitting jointly.
    }
    \label{fig:figures/contour}
\end{figure*}

We design a method to study the relationship among stellar mass, gas-phase
metallicity, SFR, and galaxy size for star-forming galaxies. We
start from the mass-metallicity relation, as shown on the left panels of
Fig.~\ref{fig:figures/mzr}. Firstly, all three observational metallicity
estimators give very similar scaling relations, where the slope is steep at the
low-mass end and gradually flattens at the high-mass end, despite the
noticeable systematics among the three estimators. Then, we employ a 4-th order
polynomial to fit the median relation, and the fitting results are shown in
black dashed lines. Finally, we calculate the standard deviation of the
gas-phase metallicity with respect to the median relation, which are
0.146/0.091/0.102 for the observational data with different metallicity
estimators and 0.158 for TNG50, respectively.

In the second step, we inspect the relationship between the residual on the
mass-metallicity relation and the SFR (not shown here), which was found to be
fitted by a linear function with the slope as $\alpha$
\citep[e.g.][]{Mannucci-10, Lara-Lopez-10,  Cresci-19, Curti-20, Curti-24,
Perez-Diaz-24}. The scatter plots of the new variable $\log M_*/{\rm M_\odot} 
- \alpha \log {\rm SFR/M_\odot yr^{-1}}$ are shown in the panels on the second
column of Fig.~\ref{fig:figures/mzr}. Again, a 4-th order polynomial was used
to fit the median relation and the standard deviations with respect to this new
scaling relation were calculated and labeled on each panel. This step can also
proceed in an equivalent way: we start from the correlation between gas-phase
metallicity and a new variable, $\log M_*/{\rm M_\odot}  - \alpha \log {\rm
SFR/M_\odot yr^{-1}}$, and fit the median relation with a 4-th order
polynomial, calculate the standard deviation of the residual. Then, we tune
$\alpha$ to minimize the standard deviation.

The standard deviation, $\sigma$, calculated on the panels in the second column
in Fig.~\ref{fig:figures/mzr} must be smaller than the those in the first
column, since one additional variable, i.e. SFR, is used to predict the
metallicity. Moreover, the decrement of $\sigma$ reflects the importance of SFR
in predicting gas-phase metallicity and it is also indicative of the physical
causation between SFR and gas-phase metallicity. Here one can see that, also
from Fig.~\ref{fig:figures/sigma}, that the decrements of the standard
deviation in gas-phase metallicity by incorporating SFR were marginal for all
there observational estimators and the TNG50 simulation.

The third step is similar to the second step, except replacing SFR with galaxy
size, $R_{\rm e}$. The results are shown in the panels on the third column of
Fig.~\ref{fig:figures/mzr}. Here one can see that the standard deviation,
$\sigma$, of gas-phase metallicity were significantly reduced by incorporating
galaxy size for both observations and the TNG50 simulation, which was more
clearly shown in Fig.~\ref{fig:figures/sigma}.

The final step is to incorporate both SFR and galaxy size, and
the results are presented in the right-most panels of
Fig.~\ref{fig:figures/mzr}. Here we can see that the standard deviations are
very close to the values on the third column (see also
Fig.~\ref{fig:figures/sigma}), which, combined with previous results, indicates
that galaxy size plays a more fundamental role in regulating the gas-phase
metal content of galaxies in their evolution than SFR.

Following all four steps above, we can obtain the standard deviations,
$\sigma$, in four cases, i.e. $(M_*)$, $(M_*, {\rm SFR})$, $(M_*, R_{\rm e})$,
and $(M_*, {\rm SFR}, R_{\rm e})$, and the results are presented on the top
panel of Fig.~\ref{fig:figures/sigma}. Here we can see that the absolute value
of standard deviation profoundly depends on the metallicity indicator.
\texttt{Scal} gives the smallest scatter and \texttt{N2S2Ha} gives the largest
scatter, while \texttt{N2O2} lies in-between. Since we only care about the role
played by SFR and galaxy size in reducing the scatter rather
than their absolute values, we normalize these scatters using the values
obtained with stellar mass alone and proceed to study the decrease by
incorporating additional galaxy properties, and the result is presented on the
middle panel of Fig.~\ref{fig:figures/sigma}. Despite the difference in the
absolute values of scatters, the normalized scatter behaves quite consistently
among three estimators: incorporating galaxy size can substantially reduce the
scatter, while SFR only plays a minor role. The more intriguing
thing is that TNG50 gives rise to qualitatively similar behavior, except that
galaxy size plays a more dominating role.

Finally, the best-fitting values of $\alpha$, $\beta$, and $(\alpha, \beta)$ in
the latter three steps are presented in Fig.~\ref{fig:figures/contour}. The
magenta triangles show the values of $\alpha$ and $\beta$ when fitted
individually with the other parameter sets to zero, and the blue crosses show
the results when fitting jointly. The background color renders the $\sigma$
value as a function of $(\alpha, \beta)$. The degeneracy between these two
parameters is obvious for both observation and simulation, and it comes from
the positive correlation between SFR and galaxy size.
Therefore, when only SFR (galaxy size) is used to fit the
residual of the mass-metallicity relation, it can leverage its correlation to
galaxy size (SFR) so that the best-fitting $\alpha$ ($\beta$)
is slightly larger than the value in the joint fitting case. This effect is
more strong at high-$z$ as we will see shortly.

\subsection{Redshift evolution in TNG50}%
\label{sub:redshift_evolution_in_tng50}

\begin{figure}
    \centering
    \includegraphics[width=1\linewidth]{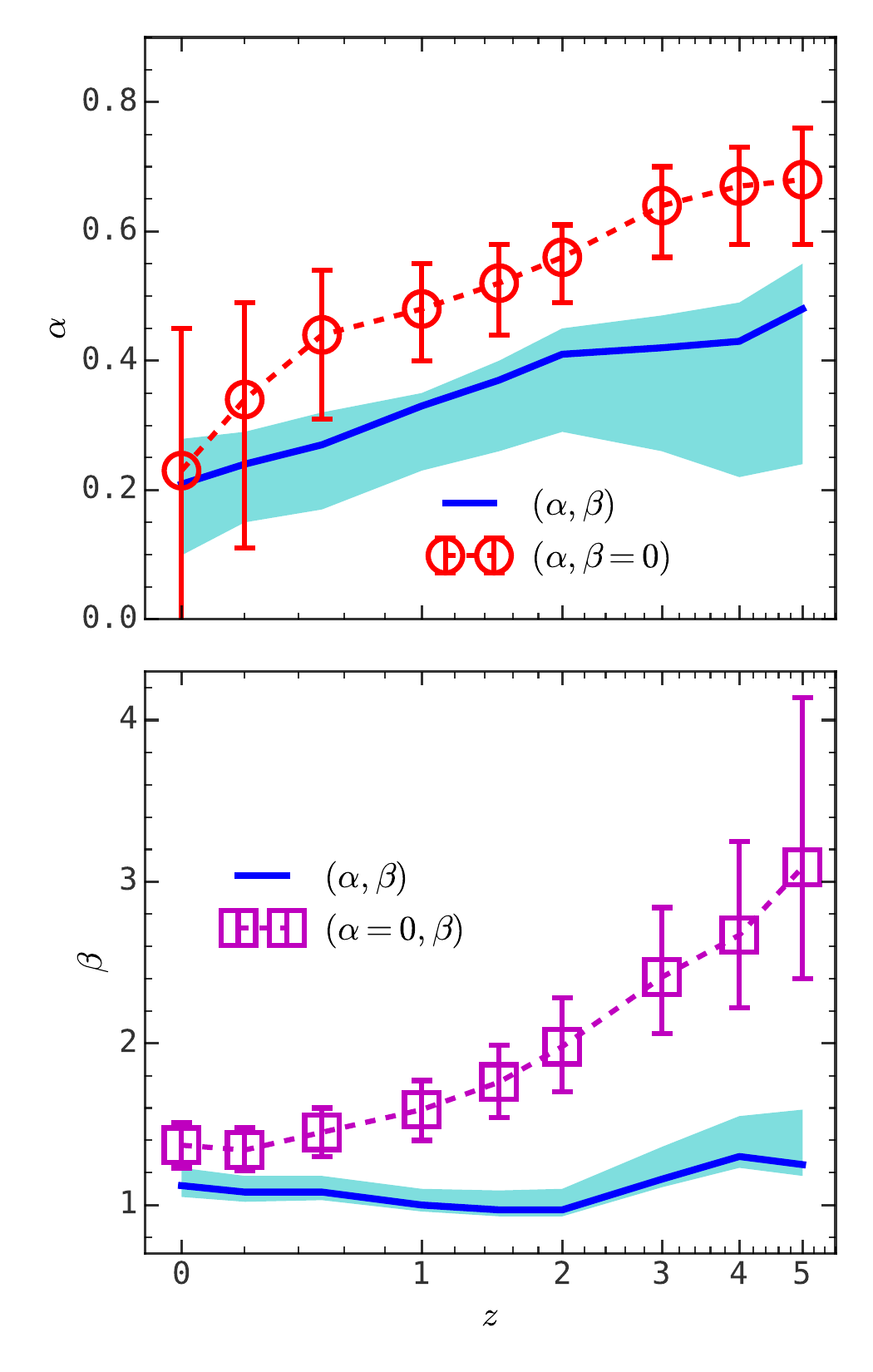}
    \caption{
        The redshift evolution of best-fitting $\alpha$ and $\beta$ in TNG50.
        The red circles in the upper panel show the best-fitting $\alpha$ when
        $\beta$ is forced to be zero, and the magenta squares show the
        best-fitting $\beta$ when $\alpha$ is forced to be zero. The blue solid
        lines and cyan shadow regions in both panels show the best-fitting
        $(\alpha, \beta)$ when both SFR and galaxy size are used to fit the
        residual of the mass-metallicity relation.
    }%
    \label{fig:figures/alpha_beta_evolution}
\end{figure}

The TNG50 simulation enables us to extend this analysis to high-$z$ universe.
As shown in the bottom panel of Fig.~\ref{fig:figures/sigma}, the galaxy size
plays a dominant role in determining the residual of the mass-metallicity
relation over the SFR up to $z\sim 5$. Meanwhile, the SFR, although in the
subdominant role, becomes increasingly important at high-$z$.

Fig.~\ref{fig:figures/alpha_beta_evolution} shows the redshift evolution of
best-fitting $\alpha$ and $\beta$ when fitting individually for SFR and galaxy
size, and $(\alpha, \beta)$ when fitting jointly. The rise of $\alpha$ with
increasing redshift in both the individual and joint fitting cases supports our
claim that the SFR becomes more strongly correlated to the residual in the
mass-metallicity relation at high-$z$.

We also note that $\beta$ increases with redshift when $\alpha$ is set to zero,
which trend, however, diminishes when both SFR and galaxy size are taken into
account. And, more importantly, $\beta$ stays close to unity in the joint
fitting case from $z=0$ to $z\approx 5$. As shown in
Appendix~\ref{sec:gravitational_potential}, the ratio between galaxy stellar
mass and galaxy size, $\log M_* - \log R_{\rm e}$, strongly correlates to the 
gravitational potential at $R_{\rm e}$, which is the work needs to be done by
moving a test particle from $R_{\rm e}$ of the galaxy to the infinity. Our
result highlights the necessity of jointly analyzing the dependence of
metallicity on SFR and size. Furthermore, the gravitational potential of
galaxies plays a primary role in determining the metallicity of galaxies in
TNG50 across the full lifetime of galaxies.

\section{Discussion and Summary}%
\label{sec:discussion_and_summary}

\subsection{Comparison with previous results}%
\label{sub:comparison_with_previous_results}

\citet{Mannucci-10} studied the relationship between the gas-phase metallicity
and SFR of star-forming galaxies, and they found a best-fitting $\alpha\approx
0.32$, which is consistent with our results here. However, they claimed that
this coefficient is independent of redshift, while TNG50 predicts that it
increases towards higher redshift.

\citet{DEugenio-18} investigated the correlation between the gas-phase metallicity and
effective radius at fixed stellar mass for star-forming galaxies in SDSS DR7.
They find these two variables in anti-correlation with the best-fitting slope
about 0.6 for galaxies with $M_*\sim 10^9\rm M_\odot$, consistent with our
results, and the slope flattens to about 0.4 for galaxies with $M_*\sim
10^{10.5}\rm M_\odot$. Similarly, \citet{sanchez-menguianoStellarMassNot2024}
exhausted 148 galaxy properties to predict the gas-phase metallicity using a
random forest regression algorithm, and they find that the compact form of
$M_*/R_{\rm e}^{0.6}$ is able to capture most of the scatter in the gas-phase
metallicity, which means that including other properties only improves the
performance marginally.

Compared with previous results, our study has several highlights. Firstly, we
employs three different metallicity indicators, aiming to the marginalize the
inherent systematics of each individual indicator. Secondly, we simultaneously
take SFR and galaxy size, which are correlated to each other
through the mass-size relation, into account to mitigate the risk that one
variable takes advantage of their mutual correlation. Finally, we apply the
same procedure to simulated galaxies, where the metallicity is calculated
directly from element abundance, and obtain qualitatively similar conclusions,
which further eliminates the risk that these results are due to observational
systematics. Moreover, it supports the modeling of metal-related processes in
the TNG simulation.

Furthermore, our study on high-$z$ galaxies in TNG50 predicts that SFR
becomes increasingly important at high-$z$. In addition, the best-fitting
$\alpha$ in increasing with redshift, while the best-fitting $\beta$ is
independent of redshift. These predictions could be tested with upcoming
high-$z$ galaxy surveys from PFS
\citep{takadaExtragalacticScienceCosmology2014} and MOONS
\citep{cirasuoloMOONSMultiObjectOptical2014, maiolinoMOONRISEMainMOONS2020}.

\subsection{Physical explanation}%
\label{sub:physical_explanation}

\begin{figure*}
    \centering
    \includegraphics[width=1\linewidth]{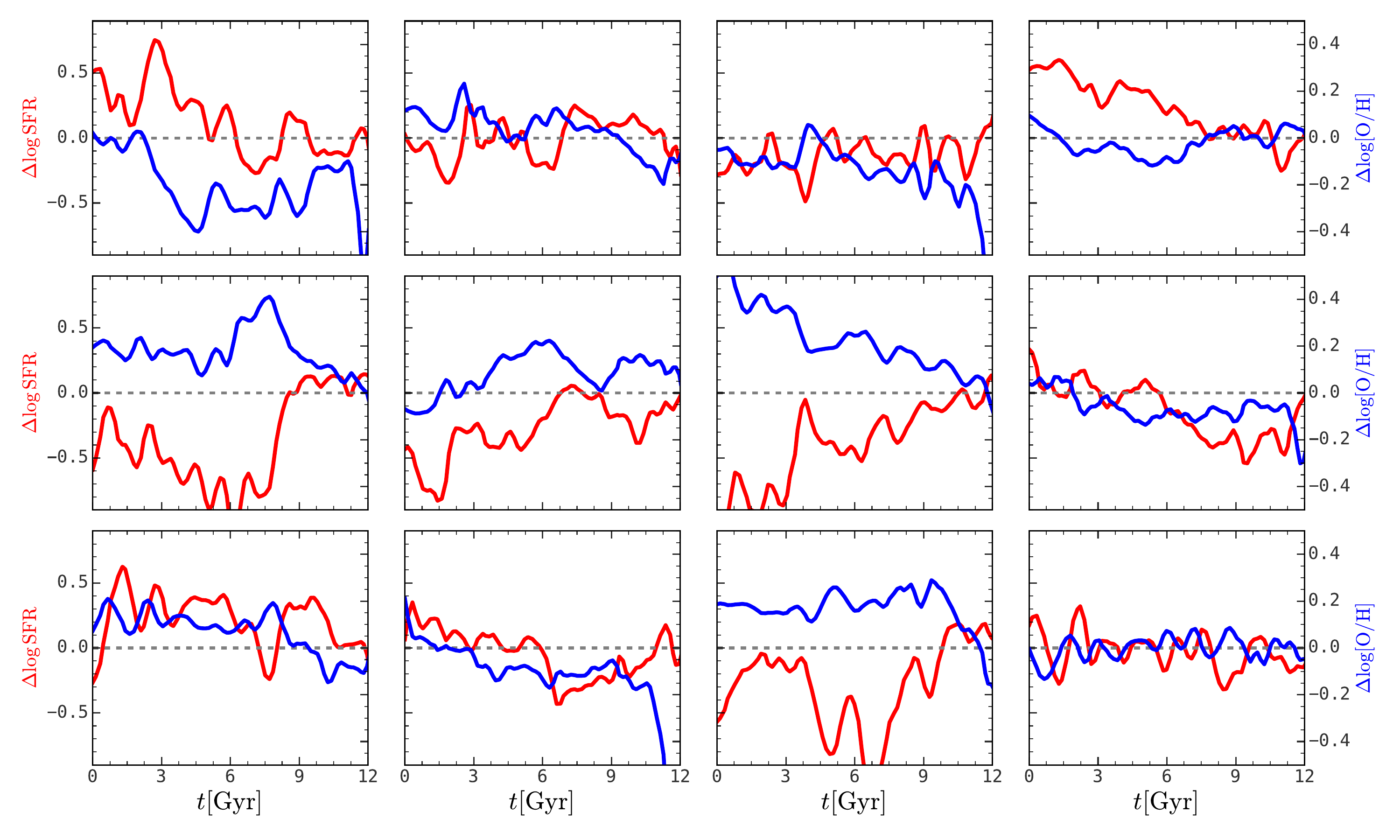}
    \caption{
        The evolution of the SFR and gas-phase metallicity for 12
        randomly selected star-forming galaxies as a function of lookback time
        in the TNG50 simulation. For clear presentation, here we show the
        deviation from the SFMS and the median MZR. This figure clearly shows
        that the SFR and the gas-phase metallicity are in anti-correlation on
        the temporal dimension.
    }%
    \label{fig:figures/evolution_example}
\end{figure*}

In order to understand the scaling relation of metallicity, we consider the
metal enrichment process under the gas regulator system
\citep[e.g.][]{Schaye-10, Dave-11, Lilly-13, Peng-14, Wang-19, Wang-21,
Wang-22}, where the instantaneous gas mass is regulated by the interplay of gas
inflow, star formation and the associated outflow. Following the work of
\cite{Wang-21}, we can write the basic continuity equations for gas and metals:
\begin{align}
    &\dot{M}_{\rm gas}(t) = \Phi(t) - (1-R)\Psi(t) - \lambda
    \Psi(t)\label{eq:1}\\
    &\dot{M}_{\rm Z}(t) = y \Psi(t) - (1-R+\lambda) Z(t)\Psi(t) + \Phi(t)
    Z_{\rm 0}\label{eq:2}
\end{align}
where $M_{\rm gas}$ is the gas mass, $M_{\rm Z}$ is the metal mass, $\Psi(t)$
is the SFR, $\Phi(t)$ is the inflow rate, $Z_0$ is the metallicity of inflowing
gas, and $y$ is the yield, i.e. the mass of metals returned to the interstellar
medium per unit mass of formed stars. $R$ denotes the fraction of mass formed
in new stars that is subsequently returned to the interstellar medium, and
$\lambda$ is the mass loading factor, i.e. the ratio between mass outflow rate
and SFR. Here we assume that the loading factor only depends on the
gravitational potential, thus $M_*/R_{\rm e}$, in our toy model.

We start from the simplest case, where $\dot{M}_{\rm gas}=0$ and $\dot{M}_{\rm
Z}=0$. Then we can obtain the metallicity of the system analytically, which can
be written as
\begin{equation}
    Z_{\rm gas} = Z_0 + y/(1-R+\lambda) \label{eq:3}
\end{equation}
We note that, even if the input inflow rate is time-varying, one still can get
the time-averaged solution in a similar form, i.e. $\langle {Z_{\rm
gas}}\rangle= Z_0 + y/(1-R+\lambda)$ \citep[see figure 4 in][]{Wang-21}.
Equation~\ref{eq:3} says that the metallicity, $Z_{\rm gas}$, is regulated by
the mass-loading factor, $\lambda$, which is directly linked to the
gravitational potential. We expect that systems with deeper gravitational
potential, i.e. smaller galaxy sizes, are more resistant to the stellar
feedback process, thus have lower outflow rate and lower mass loading factor,
at given SFR. Consequently, these systems are more metal-rich.
This explains the correlations we saw previously in observation and simulation,
where galaxies with smaller sizes are more metal-rich at given stellar mass.

In addition, inputting time-varying inflow rate into Equation~\ref{eq:1} and
\ref{eq:2}, \cite{Wang-21} found a clear negative correlation between SFR and
$Z_{\rm gas}$ for individual gas regulator system. It is not possible to trace
the time-variation for individual galaxies from the observation, while the
simulations provide this important information. We therefore examine the
correlation of $\Delta \log {\rm SFR}$ and $\Delta \log ({\rm O/H})$ as a
function of redshift for 12 TNG50 galaxies (randomly selected), shown in Figure
\ref{fig:figures/evolution_example}. The $\Delta \log {\rm SFR}$ and $\Delta
\log ({\rm O/H})$ are defined as the offsets from the SFMS and MZ relation
established at corresponding redshifts. Fig~\ref{fig:figures/evolution_example}
clearly shows negative correlations between $\Delta \log {\rm SFR}$ and $\Delta
\log ({\rm O/H})$ for individual galaxies, indicating that the dependence of
metallicity on SFR is indeed from the time-variation of individual galaxies, at
least for TNG50. Compared to gravitational potential, SFR plays a secondary
role in determining the metallicity, driven by the time-varying inflow rate.

Nonetheless, the cautious reader may have noticed some positive correlations
between $\Delta \log {\rm SFR}$ and $\Delta \log \rm [O/H]$, like the
second panel on the bottom row. According to the toy model proposed in
\citet{Wang-21}, a time-varying SFR could only induce a negative
correlation between SFR and the gas-phase metallicity, while a time-varying
${\rm SFE}\equiv {\rm SFR}/M_{\rm gas}$ could cause a positive correlation.
So we inspect the SFE histories of these galaxies and indeed find an upturn
of SFE for this particular galaxy (not shown here). Therefore, the rise of
SFR for this galaxy from $z\approx 1$ is caused by an enhanced SFE instead of
increasing inflow gas, so that the gas-phase metallicity is not diluted but,
instead, enriched by the metal produced during the star formation process.
Consequently, both SFR and the gas-phase metallicity arise and produce a
positive correlation. Apparently, SFE also plays an important role in
regulating the gas-phase metallicity of galaxies. However, calculating SFE
requires estimating the total gas mass, which is quite expensive in observation
and limited to a small sample of galaxies to date. In addition, the averaged
SFE of a whole galaxy does not vary too much temporally \citep{Wang-21}, so
that we could ignore it in most cases.

\subsection{Summary}%
\label{sub:summary}

The gas-phase metallicity of galaxies contains abundant information about
astrophysical processes in galaxy formation and evolution. A common practice to
decode this information is to first establish scaling relations with other
galaxy properties and, then, compare them with galaxy formation models. Beyond the
mass-metallicity relation, people start to pursue the secondary galaxy property
that possesses the strongest correlation to the gas-phase metallicity, and a
debate between SFR and galaxy size arises. In this work, we use both
cutting-edge observations and state-of-the-art simulations examine the roles
played by these galaxy properties in determining the gas-phase metallicity. Our
findings are summarized as follows.

\begin{enumerate}

    \item We find that, based on the mass-metallicity relation, the inclusion
        of galaxy size, $\log M_* - \beta\log R_{\rm e}$, can significantly
        reduce the uncertainty in predicting the gas-phase metallicity,
        compared with using stellar mass alone, for star-forming galaxies in
        the MaNGA survey with three different metallicity observational
        indicators, despite their inherent systematics. Meanwhile, SFR only
        plays a subdominant role for all three metallicity indicators. Similar
        conclusions are drawn in the TNG50 simulation.

    \item The best-fitting coefficients for SFR and galaxy size
        are $\alpha\approx 0.2$ and $\beta\approx 0.6$, respectively, for three
        different metallicity indicators for the MaNGA survey, while TNG50
        gives $\alpha\approx 0.5$ and $\beta\approx 1.0$.

    \item Through performing similar analyses at different redshifts in TNG50,
        we find that SFR plays an increasingly important role in
        predicting metallicity at high-$z$, and the best-fitting value of
        $\alpha$ also increases with redshift. On the other side, the
        best-fitting value of $\beta$ is always $\approx 1$ from $z\approx 0$
        to $z\approx 5$.

    \item Based on the statistical analysis of the relationship among gas-phase
        metallicity, galaxy size, and SFR, we speculate that SFR modulates
        metallicity on the temporal dimension, synchronized to the gas inflow
        process, while galaxy size regulates metallicity on the spatial
        dimension through affecting the gravitational potential and the mass
        loading factor.

\end{enumerate}

Our analysis here mainly focuses on statistical analysis in our local Universe.
Nevertheless, with the advent of high-$z$ galaxy surveys, we expect to
investigate the statistical relationship between metallicity and other galaxy
properties at cosmic noon to understand the underlying astrophysical process,
and even directly inspect the metal exchange between galaxies and their
surrounding medium \citep{zhangInspiralingStreamsEnriched2023}, which is more
prevalent in early Universe.

\section*{Acknowledgements} The authors thank the anonymous referee for their
helpful comments that improved the quality of this paper. This work is
supported by the National Science Foundation of China (NSFC) Grant No.
12233008, 12125301, 12192220, 12192222, the National Key R\&D Program of China
(2023YFA1608100), and the science research grants from the China Manned Space
Project with NO. CMS-CSST-2021-A07. The authors gratefully acknowledge the
support of Cyrus Chun Ying Tang Foundations.

\bibliographystyle{aasjournal}
\bibliography{bibtex.bib}

\appendix

\section{Gravitational potential}%
\label{sec:gravitational_potential}

\begin{figure}
    \centering
    \includegraphics[width=0.5\linewidth]{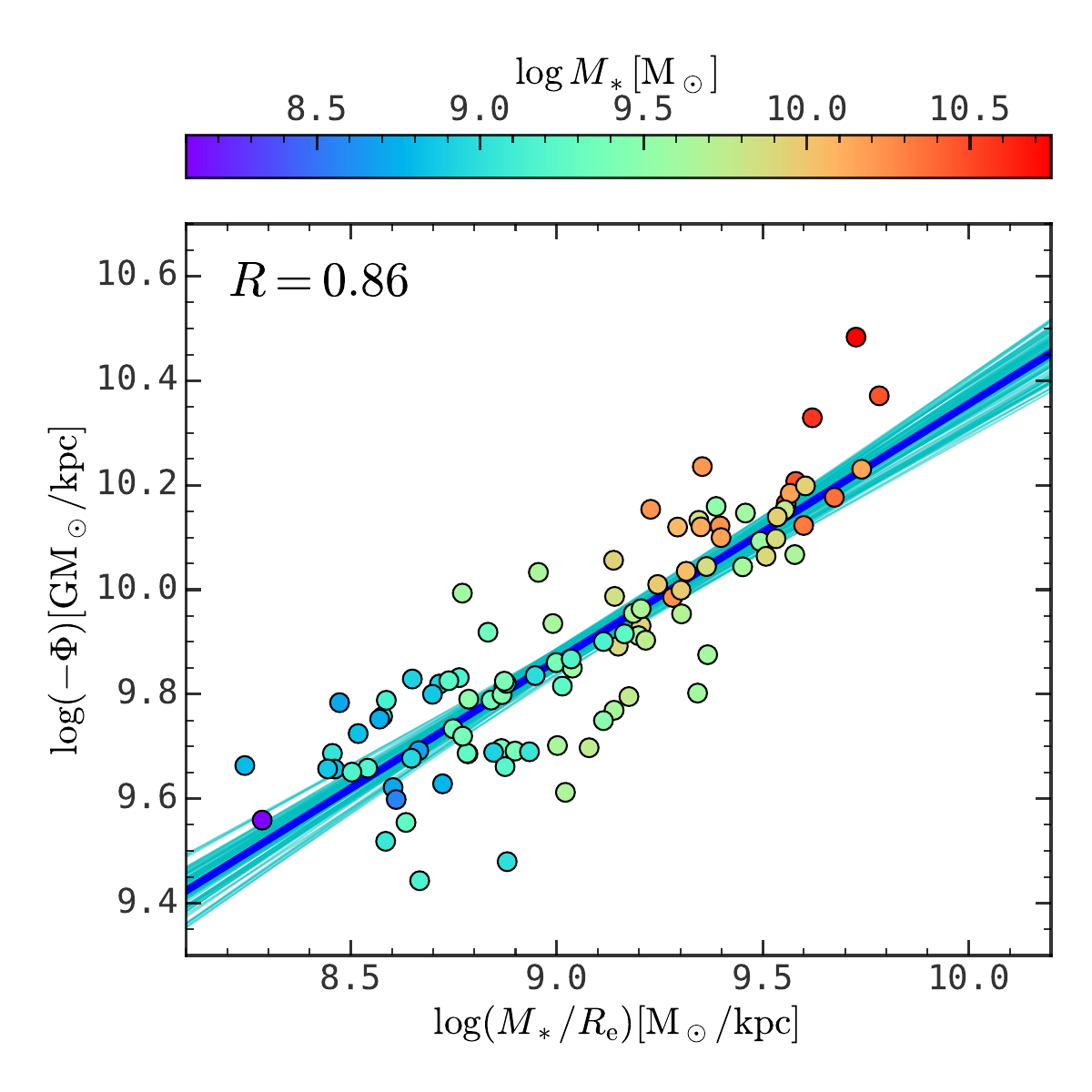}
    \caption{
        The gravitational potential versus the $M_*/R_{\rm e}$ for a randomly
        selected a sample of 100 galaxies.  The blue line shows the linear
        fitting for the two quantities with the correlation coefficient of
        0.86, indicating strong correlations between the
    two. }%
    \label{fig:figures/potential}
\end{figure}

Here we examine whether the $M_*/R_{\rm e}$ can be a good indicator of the 
gravitational potential of the whole system, including both baryonic and dark
matter particles. We randomly select 100 galaxies that spans a wide range of
stellar mass for this test.

We calculate the gravitational potential at the radius of $R_{\rm e}$ in the
following way.  For simplicity, we calculate the mean potential of a sphere
with the radius of $R_{\rm e}$ centered at galactic center.  In this case, the
mean potential at this sphere for a given particle (with mass of {$M_{\rm
par}$}) within the sphere can be expressed as  $-GM_{\rm par}/R_{\rm e}$, and
this value changes to be $-GM_{\rm par}/r$ for a particle out of the sphere,
where $r$ is the distance to the galaxy center. Then we sum over all the
particles to obtain the mean gravitational potential of individual galaxies.

Fig. \ref{fig:figures/potential} shows the correlation between the
gravitational potential at $R_{\rm e}$ calculated above and the $M_*/R_{\rm
e}$. Interestingly, we do find the two show a tight relation, with a
correlation coefficient of 0.86.  This confirms that $M_*/R_{\rm e}$ indeed can
be a good tracer as the overall gravitational potential of galaxies.




\end{document}